\documentstyle[prl,aps,epsf]{revtex}
\draft
\newcommand{\be}{\begin{equation}}
\newcommand{\bea}{\begin{eqnarray}}
\newcommand{\ee}{\end{equation}}
\newcommand{\eea}{\end{eqnarray}}

\begin{document}

\title{The Pinch Technique at Two Loops}

\author{Joannis Papavassiliou}

\address{Departamento de F\'{\i}sica Te\'orica, Univ. Valencia \\
E-46100 Burjassot (Valencia), Spain}

\maketitle

\begin{abstract} 
It is shown 
that the fundamental properties of gauge-independence, gauge-invariance,
unitarity, and analyticity
of the $S$-matrix lead to the unambiguous  
generalization of 
the pinch technique algorithm to two loops.
  
\medskip

\noindent
PACS numbers: 11.15.-q, 12.38.Bx, 14.70.Dj, 11.55.Fv ~~~~FTUV-99-12-14
~~~~ \\
e-mail: Joannis.Papavassiliou@cern.ch;@uv.es
\end{abstract}

\vspace{1.cm}

A variety of important physical problems cannot be addressed within
the framework of fixed-order perturbation theory, the most widely used
calculational scheme in the continuum.  This is often the case within
Quantum Chromodynamics (QCD), when large disparities in the physical
scales involved result in a complicated interplay between perturbative
and non-perturbative effects.  Similar limitations appear when
physical kinematic singularities, such as resonances, render the
perturbative expansion divergent at any finite order, or when
perturbatively exact symmetries prohibit the appearance of certain
phenomena, such as chiral symmetry breaking or gluon mass generation.
In such cases one often resorts to various reorganizations of the
perturbative expansion inspired from scalar field theories or Quantum
Electrodynamics (QED), supplemented by a number of auxiliary physical
principles.  When studying the interface between perturbative and
non-perturbative QCD for example, one finds it advantageous to use
concepts familiar from QED, such as the effective charge, in
conjunction with dispersive techniques and analyticity properties of
the $S$-matrix \cite{DMW}.  In addition, in the field of renormalon
calculus, one studies the onset of non-perturbative effects from the
behaviour near the QCD mass-scale of judiciously selected infinite
sub-sets of the perturbative series
\cite{MB}.
Similarly, the extension of the Breit-Wigner formalism 
to the electro-weak sector of the 
Standard Model necessitates a non-trivial rearrangement of the
perturbative expansion  \cite{PP}; 
an analogous task must be undertaken when studying various aspects of
finite temperature QCD  \cite{Tqcd}, 
as well as mass generation, both in
3-dimensional field-theories \cite{3d}
and in QCD \cite{JMC}, as a prelude to a systematic truncation 
scheme for the Schwinger-Dyson series.

One of the main difficulties encountered when dealing with the
problems mentioned above is the fact that several physical properties,
which are automatically preserved in fixed-order perturbative
calculations by virtue of powerful field-theoretical principles, may
be easily compromised when rearrangements of the perturbative series,
such as resummations, are carried out.  These complications may in
turn be traced down to the fact that in non-Abelian gauge theories
individual off-shell Green's functions ($n$-point functions) are in
general unphysical.
 
It turns out that this last problem can be circumvented   
by resorting to the method known as the pinch technique (PT)
\cite{JMC,JMCJP}.
The  PT 
reorganises     systematically    a given   physical   amplitude  into
sub-amplitudes,   which have   the   same   kinematic  properties   as
conventional $n$-point functions, (propagators, vertices, boxes)
\cite{KL},
but, in addition, are endowed with desirable physical properties.
Most importantly, at one-loop order
(i) are independent of  the  gauge-fixing
parameter; (ii)  satisfy naive,
(ghost-free) tree-level    Ward  identities, instead    of   the usual
Slavnov-Taylor identities.
(iii) contain only physical thresholds and satisfy very special 
unitarity relations  \cite{PP,PRW}
(iv)  coincide with the conventional $n$-point functions when the
latter are computed in the background field method Feynman gauge 
(BFMFG)\cite{BFMPT}.
These properties are realized {\it diagrammatically}
by exploiting  the elementary 
Ward identities of  the  theory  in  order to enforce  crucial
cancellations\cite{FL}, and make manifest intrinsic properties 
of the $S$-matrix, which are usually concealed by the quantization
procedure.

The important question  which  arises is  whether  the
PT algorithm may be extended beyond one-loop, leading to  
 the systematic replication of the  aforementioned    special
properties of  the PT effective $n$-point functions  to higher orders
\cite{BFMPT2F}.
In this  Letter we will show that the PT can 
be generalized to two loops   
by resorting {\it exactly} 
to the same physical and field-theoretical principles 
as at one-loop. 

We start by briefly reviewing the one-loop case.
Consider the $S$-matrix element for the  
quark ($u$)-antiquark ($\bar{u}$) scattering process 
$u(P)\bar{u}(P')\to u(Q)\bar{u}(Q')$ in QCD;
we set $q= P'-P= Q'-Q$, and $s=q^2$ is  
the square of the momentum transfer.  
It is convenient to work
in the renormalizable Feynman gauge (RFG); this constitutes no
loss of generality,
since the full $S$-matrix is independent of the gauge-fixing parameter
and gauge-fixing scheme. 
One first decomposes the elementary tree-gluon vertex 
$\Gamma_{\alpha\mu\nu}^{(0)}(q,p_1,p_2)$ as follows
\cite{JMCJP}:
\bea
\Gamma_{\alpha\mu\nu}^{(0)}(q,p_1,p_2) &=&
[(p_1-p_2)_{\alpha} g_{\mu\nu} + 2q_{\nu}g_{\alpha\mu} 
- 2q_{\mu}g_{\alpha\nu}] + 
 [p_{2\nu} g_{\alpha\mu} - p_{1\mu}g_{\alpha\nu}]
\nonumber\\  
 &=& \Gamma_{F\alpha\mu\nu}^{(0)}(q,p_1,p_2) + 
\Gamma_{P\alpha\mu\nu}^{(0)}(q,p_1,p_2) ~.
\label{decomp}
\eea
This decomposition assigns a special role to the $q$-leg,
and allows $\Gamma_{F\alpha\mu\nu}^{(0)}$ 
to satisfy the Ward identity
\be 
q^{\alpha} \Gamma_{F\alpha\mu\nu}^{(0)}(q,p_1,p_2) = 
(p_2^2 - p_1^2) g_{\mu\nu}
\label{WI2B}
\ee
where the right-hand-side is the difference of two-inverse
propagators in the Feynman gauge, and vanishes on shell,
i.e. $p_1^2=p_2^2=0$.
Notice that the
first term in $\Gamma_{F\alpha\mu\nu}^{(0)}$ is a convective
vertex, 
whereas the
other two terms originate from gluon spin or magnetic moment. 
$\Gamma_{F\alpha\mu\nu}^{(0)}(q,p_1,p_2)$
coincides with the BFMFG bare vertex involving one
background ($q$) and two quantum ($p_1$,$p_2$) gluons
\cite{BFMPT}. 

We then carry out the above decomposition on the
three-gluon vertex appearing
inside the non-Abelian graph
contributing to the one-loop quark-gluon vertex \cite{JMCJP} .
The result of this is two-fold: First,
the action of the longitudinal momenta 
$p_{1}^{\mu}=-k^{\mu}$, $p_{2}^{\nu}=(k-q)^{\nu}$
on the bare quark-gluon vertices 
$\Gamma_{\mu}^{(0)}$ and $\Gamma_{\nu}^{(0)}$, 
respectively,
triggers the elementary Ward identity of the form
$\not\!  k = (\not\! k + \not\! Q -m) - (\not\! Q -m)$.
The first term gives rise to the
pinch contribution $V_{P\alpha\sigma}^{(1)}(q)$ given by
$V_{P\alpha\sigma}^{(1)}(q) = 2 g^2 C_A 
\int [dk][k^2 (k+q)^{2}]^{-1} ~g_{\alpha\sigma}$, 
where $g$ is the gauge coupling, $C_A$ is the
Casimir eigenvalue of the adjoint representation, and 
$ [dk] = \mu^{2\epsilon}\frac{d^d k}{(2\pi)^d}$, with 
$\mu$ the 't Hooft mass; the second term vanishes on-shell.
Second, the part of the graph containing $\Gamma_{F\alpha\mu\nu}^{(0)}$
together with its Abelian-like counterpart defines the PT  
one-loop quark-gluon vertex 
$\widehat{\Gamma}_{\alpha}^{(1)}(Q,Q')$,
which satisfies the QED-like Ward identity 
$q^{\alpha}\widehat{\Gamma}_{\alpha}^{(1)}(Q,Q')=
\widehat{\Sigma}^{(1)}(Q)-\widehat{\Sigma}^{(1)}(Q')$,
where $\widehat{\Sigma}^{(1)}$ is the PT one-loop
quark self-energy.
The propagator-like parts extracted from the vertex
are cast into the form of a genuine self-energy by setting 
$\Pi_{P\alpha\beta}^{(1)}(q) =  
V_{P}^{(1)}(q)_{\alpha\sigma}t^{\sigma}_{\beta}(q) $, 
where $t_{\mu\nu}(q) = q^2 g_{\mu\nu} - q_{\mu}q_{\nu}$;
thus, the resulting one-loop PT self-energy reads
$\widehat{\Pi}_{\alpha\beta}^{(1)}(q) =  \Pi_{\alpha\beta}^{(1)}(q)
+ \Pi_{P\alpha\beta}^{(1)}(q)$.
Carrying out the one-loop integrations one finds \cite{JMC}
that the prefactor in front of the logarithm of 
$\widehat{\Pi}_{\alpha\beta}^{(1)}(q)$ is $(11/3)C_A$, i.e.
the coefficient of the one-loop
$\beta$ function for quark-less QCD.

For the two-loop case, one considers the
two-loop $S$-matrix element for the
aforementioned process $u\bar{u}\to u\bar{u}$ in the RFG,
and focusses on the two-loop quark-gluon vertex
$\Gamma_{\alpha}^{(2)}(Q,Q')$.
The Feynman graphs contributing to $\Gamma_{\alpha}^{(2)}(Q,Q')$
can be classified into two sets.
(a) those containing an ``external'' three-gluon vertex 
i.e. a three-gluon vertex where the momentum $q$ is incoming
(Fig.1).
(b) those which do not have an ``external'' three-gluon vertex.
This latter set contains either graphs with no three gluon vertices
(abelian-like), or graphs with
three-gluon vertices whose all three
legs are irrigated by virtual momenta, i.e. $q$ never enters  
alone into any of the legs. 
Carrying out the decomposition of Eq. (\ref{decomp})
to the external three-gluon vertex   
of all graphs belonging to set (a), leaving {\it all} their other
vertices unchanged \cite{NJW3}, 
the following
situation emerges: 
\be
\Gamma_{\alpha}^{(2)}(Q,Q')
=\widehat{\Gamma}_{\alpha}^{(2)}(Q,Q')+
\frac{1}{2}V_{P\alpha}^{(2)\sigma}(q)\Gamma_{\sigma}^{(0)} + 
\frac{1}{2}\Pi_{P\alpha}^{(1)\beta}(q) (\frac{-i}{q^2})
\widehat{\Gamma}_{\beta}^{(1)}(Q,Q') ~,
\label{2LV}
\ee
with
\bea
V_{P\alpha\sigma}^{(2)}(q) &=&
-I_1 \Bigg[k_{\sigma}g_{\alpha\rho}+  
\Gamma_{\rho\sigma\alpha}^{(0)}(-k,-\ell,k+\ell)\Bigg] (\ell-q)^{\rho}
+ (2 I_{2} + I_{3}) g_{\alpha\sigma} 
\nonumber\\
&&I_4 
[\Gamma_{\alpha\lambda\rho}^{(0)}(\ell,k,-k-\ell)
\Gamma_{\sigma}^{(0)\lambda\rho}(\ell,k,-k-\ell) - 
2k_{\alpha}(k+\ell)_{\sigma}] ~,
\label{2LVP}
\eea
where $I_{1} = I_{0} (k+\ell)^{-2} (k+\ell-q)^{-2}$,
$I_{2} = I_{0}(k+q)^{-2}$, $I_{3} = I_{0}(k+\ell)^{-2}$,
$I_{4} = I_{0} \ell^{-2} (k+\ell)^{-2}$, with
$iI_{0} = g^4 C_A^2 [\ell^2 (\ell-q)^2 k^2]^{-1}$, 
and 
the two-loop integration prefactor
$(\mu^{2\epsilon})^2
\int\int \frac{d^d k}{(2\pi)^d} \frac{d^d \ell}{(2\pi)^d}$
has been suppressed. 
$\widehat{\Gamma}_{\alpha}^{(2)}(Q,Q')$ is the two-loop
BFMFG quark-gluon vertex, $V_{P\alpha\sigma}^{(2)}(q)$
the propagator-like part, and the third term on the 
right-hand side is the necessary contribution for converting
the one-particle reducible part of the two-loop $S$-matrix element
 ${\Gamma}_{\alpha}^{(0)}
(\frac{-i}{q^2})\Pi_{\alpha\beta}^{(1)}(q) (\frac{-i}{q^2})
{\Gamma}_{\beta}^{(1)}(Q,Q')$ into 
${\Gamma}_{\alpha}^{(0)}(\frac{-i}{q^2}) 
\widehat\Pi_{\alpha\beta}^{(1)}(q) (\frac{-i}{q^2})
\widehat{\Gamma}_{\beta}^{(1)}(Q,Q')$.
Eq.(\ref{2LV}) 
is a non-trivial result, since there is no a-priori
reason why the implementation of the decomposition 
of Eq. (\ref{decomp}) 
should only give rise to terms 
which can be interpreted in the way described above.  
In fact, individual diagrams, or even natural sub-sets of diagrams
such as the one-loop three-gluon vertex nested inside the
two-loop quark-gluon vertex, give in general rise to contributions 
which do
{\it not} belong to any of 
the terms on the right-hand side of Eq. (\ref{2LV}) . 
It is only after all
terms have been considered that the aforementioned crucial 
cancellations become possible. Finally, 
the counterterms
of $\Gamma_{\alpha}^{(2)}(Q,Q')$ must be correctly accounted
for \cite{P2}.   
$\widehat{\Gamma}_{\alpha}^{(2)}(Q,Q')$
satisfies the QED-like
Ward identity 
$q^{\alpha}\widehat{\Gamma}_{\alpha}^{(2)}(Q,Q')=
\widehat{\Sigma}^{(2)}(Q)-\widehat{\Sigma}^{(2)}(Q')$,
where
 $\widehat{\Sigma}^{(2)}$ is the two-loop PT quark-self-energy.
$\widehat{\Sigma}^{(2)}$ is
identical to the conventional  $\Sigma^{(2)}$
in the RFG (and the BFMFG),
exactly as happens at one-loop.

To construct the two-loop PT gluon self-energy 
$\widehat{\Pi}_{\alpha\beta}^{(2)}(q)$, one 
must append to the
conventional two-loop self-energy  $\Pi_{\alpha\beta}^{(2)}(q)$
the 
term 
$\Pi_{P\alpha\beta}^{(2)}(q) =  
V_{P\alpha\sigma}^{(2)}(q)t^{\sigma}_{\beta}(q) $
{\it together} with the term 
$iR_{P\alpha\beta}^{(2)}(q)  = 
\Pi_{\alpha\beta}^{(1)}(q)V^{(1)}_{P}(q) + 
\frac{3}{4}V^{(1)\sigma}_{P\alpha}(q)
\Pi_{P\sigma\beta}^{(1)}(q)
$
originating from converting a string
of two conventional one-loop self-energies into a string
of two one-loop PT self-energies \cite{PP}. 
One can show by means of a 
diagram-by-diagram mapping
that the resulting 
$\widehat{\Pi}_{\alpha\beta}^{(2)}(q)$
is {\it exactly} identical to the corresponding two-loop self-energy
of the BFMFG, and that this correspondence persists
after renormalization \cite{P2}. Notice
that the
presence of the term $R^{(2)}_{P\alpha\beta}(q)$ is crucial for
the entire construction, and constitutes 
a non-trivial consistency check
of the resummation mechanism first proposed in \cite{PP}.
An immediate 
consequence of the above correspondence is that the 
coefficient in front of the leading logarithm of 
$\widehat{\Pi}_{\alpha\beta}^{(1)}(q)$ 
is precisely the
coefficient of the two-loop 
quark-less QCD $\beta$ function \cite{Abbott}, 
namely $(34/3)C_A^2$ \cite{WEC}. 
As a result, one may extend to two-loops
the one-loop construction of a renormalization-group-invariant
effective charge presented in \cite{NJW2}, leading
to the unambiguous identification of the conformally-(in)variant
subsets of QCD graphs\cite{BLM}.
Finally we note that, exactly as happens at one-loop,
the two-loop PT box-graphs are simply the conventional ones in the 
RFG (and are equal to the ones in the BFMFG).

As has been explained in detail in \cite{PP,PRW}, the
one-loop PT $n$-point functions
satisfy the optical theorem {\it individually}.
To verify that 
one starts with the tree-level process 
$u(P)\bar{u}(P') \to g(p_1) + g(p_2)$, whose  
$S$-matrix element we denote by 
${\cal T}_{\mu\nu}$; 
then, one considers  
the quantity
${\cal T}_{\mu\nu}P^{\mu\mu'}(p_1) P^{\nu\nu'}(p_2){\cal T}_{\mu'\nu'}$,
where
$ P_{\mu\nu}(p,\eta )\ =\ -g_{\mu\nu}+ (\eta_{\mu}p_{\nu}
+\eta_{\nu}p_{\mu} )/{\eta p} + 
\eta^2 p_{\mu}p_{\nu}/{(\eta p)}^2$, with $\eta$ 
an arbitrary four-vector. One proceeds
by first eliminating the  
$\Gamma_{P\alpha\mu\nu}^{(0)}(q,p_1,p_2)$ 
part of $\Gamma_{\alpha\mu\nu}^{(0)}(q,p_1,p_2)$,
which vanishes when contracted with the term
$P^{\mu\mu'}(p_1) P^{\nu\nu'}(p_2)$. 
Then, the longitudinal parts of the  
$P_{\mu\mu'}(p_1)$ and $P_{\nu\nu'}(p_2)$ trigger
a fundamental cancellation \cite{PP,PRW}
involving the $s$- and $t$- channel graphs,
which is a consequence of the underlying BRS symmetry
\cite{BRS}. 
Specifically, the action of $p_{1\mu}$ on the 
$\Gamma_{F\alpha\mu\nu}^{(0)}$ gives 
 \be
p_{1}^{\mu}\Gamma_{F\alpha\mu\nu}^{(0)}(q,p_1,p_2)=
t_{\alpha\nu}(q) + (p_1^2-p_2^2)g_{\alpha\nu} + 
(p_2-p_1)_{\alpha}p_{2\nu} ~;
\label{WI1L}
\ee
the first term on the right-hand side
cancels against an analogous contribution
from the $t$-channel graph, 
whereas the second term vanishes 
for on-shell gluons. Finally, the  
term proportional to $p_{2\nu}$ is such that 
(i) all dependence on $\eta$ vanishes, and  (ii) 
a residual contribution emerges, which must be added
to the parts stemming from the $g_{\mu\mu'}g_{\nu\nu'}$
part of the calculation.
Then one simply defines 
self-energy/vertex/box-like sub-amplitudes according
to the dependence on $s=(p_1+p_2)^2$ and $t=(P-p_1)^2$, as in 
a scalar theory, or QED. The emerging structures correspond
to the imaginary parts of the one-loop PT 
effective Green's functions, as one can readily verify
by employing the Cutkosky rules; in fact the residual pieces
mentioned at step (ii) above correspond precisely 
to the Cutkosky cuts of the one-loop ghost diagrams.
The one-loop PT structures may be 
reconstructed directly from this tree-level 
calculation, without resorting to 
an intermediate diagrammatic interpretation,
by means of
appropriately subtracted dispersion relations.

The same procedure must be followed at two-loops;
the only difference is that one must now combine
contributions from both the one-loop $S$-matrix element for
the process $u(P)\bar{u}(P') \to g(p_1) + g(p_2)$ 
and the tree-level $S$-matrix element for the process 
$u(P)\bar{u}(P') \to g(p_1) + g(p_2)+ g(p_3) $ . 
The non-trivial point is that the
one-loop $S$-matrix element must be cast 
into its PT form (as shown in Fig 2a.)
{\it before} any further manipulations take place.
Notice that the same procedure 
which leads to the appearance of $\widehat{\Pi}(q)$ \cite{NJW1}
leads also to the conversion of the
conventional one-loop three-gluon vertex  
 $\Gamma_{\,\alpha\mu\nu}^{(1)}(q,p_1,p_2)$
into $\Gamma_{F\,\alpha\mu\nu}^{(1)}(q,p_1,p_2)$, which
is the BFMFG one-loop three-gluon vertex
with one background ($q$) and two quantum ($p_1$, $p_2$) 
\cite{P2}. 
It is straightforward to show that 
$\Gamma_{F\,\alpha\mu\nu}^{(1)}(q,p_1,p_2)$
satisfies the following Ward identity
\be
q^{\alpha} \Gamma_{F\,\alpha\mu\nu}^{(1)}(q,p_1,p_2)
= \Pi_{\mu\nu}^{(1)}(p_1) - \Pi_{\mu\nu}^{(1)}(p_2) , 
\ee
which is the exact one-loop analogue of 
the tree-level Ward identity
of Eq (\ref{WI2B}); indeed the right-hand side is the
difference of two one-loop self-energies computed in the RFG.
In order to extend to the next order the dispersive construction
outlined above, one
needs the following Ward identity

\be
p_{1}^{\mu} \Gamma_{F\,\alpha\mu\nu}^{(1)} =
i \widehat\Pi_{\alpha\nu}^{(1)}(q) - i \Pi_{\alpha\nu}^{(1)}(p_2)
+ \lambda^{(1)}_{\nu\sigma}t^{\sigma}_{\alpha}(q)
+ s^{(1)}_{\alpha}p_{2\nu}
\label{WI2L}
\ee
with
\bea
\lambda^{(1)}_{\nu\sigma} &=& 
J_3 \Bigg [(k-p_1)^{\rho}
\Gamma_{\nu\rho\sigma}^{(0)}(p_2,k,-k-p_2)
-(k+p_2)_{\nu}k_{\sigma}\Bigg] - i \Bigg [2 B(q)+B(p_1)\Bigg] g_{\nu\sigma} 
\nonumber\\
s^{(1)}_{\alpha} &=& 
J_3 \Bigg [ 
p_{2}^{\sigma} k^{\rho} \Gamma^{(0)}_{F\alpha\sigma\rho}(q,k+p_2,-k+p_1)
- p_2 \cdot (k-p_1) (2k+p_2-p_1)_{\alpha}\Bigg] 
+ \bigg(\frac{1}{8}\bigg)\Bigg[B(p_1)+B(p_2)\Bigg] q_{\alpha} ,
\label{2LGH2}
\eea  
$J \equiv \frac{1}{2} ig^2 C_A [k^2 (k-p_1)^2 (k+p_2)^2]^{-1}$ 
and $B(p)\equiv g^2 C_A \int [dk][k^2 (k+q)^{2}]^{-1}$.
 Eq. (\ref{WI2L})
is the one-loop analogue of Eq. (\ref{WI1L}) . 
The one-loop version 
of the fundamental BRS-driven cancellation will then be
implemented; for instance, the first term on the 
right-hand side of 
Eq. (\ref{WI2L}) 
will cancel against analogous contributions 
from the graph of Fig. 2$a_2$,
whereas all remaining  terms proportional to $t_{\sigma\alpha}(q)$
will cancel against contributions from the $t$-channel graphs of 
Fig. 2$a_3$

The same construction
 must then be repeated for the tree-level process
$u\bar{u} \to ggg$, whose tree-level $S$-matrix element 
we denote by ${\cal T}_{\mu\nu\rho}$ ; 
again, the $s$-channel graphs (Fig. 2b) must be 
rewritten
in such a way that when contracted
with $q$ only terms proportional to $p_i^2$ emerge,
but no transverse pieces, exactly as in Eq.(\ref{WI2B}).
This is accomplished by simply  
carrying out the decomposition of Eq.(\ref{decomp}) only to 
the vertices where $q$ is entering;
then the contributions originating from the $\Gamma_{P\alpha\mu\nu}^{(0)}$
parts eventually vanish
when contracted with the polarization tensors
$P^{\mu\mu'}(p_1) P^{\nu\nu'}(p_2) P^{\rho\rho'}(p_3)$. 
Acting with the longitudinal parts of the polarization tensors on the 
${\cal T}_{\mu\nu\rho}$
one must first carry out 
the corresponding BRS $s-t$ channel cancellation, and 
pick up {\it automatically} 
the correct ghost parts. 
Notice in particular that
this procedure gives rise
to the ghost structure given in Fig.3c of \cite{Abbott}, 
which has only three-particle Cutkosky cuts, 
and does not exist in the conventional formulation.

Adding the $s$-channel terms together the  total propagator-like
part emerges; it is 
proportional to $(34/3)C_A^2q^2$, as it should.
Notice that the result is 
infrared finite,
by virtue of crucial cancellations between the 
the one-loop $u\bar{u}\to gg$ and 
the tree-level $u\bar{u}\to ggg$ cross-sections. 
The most direct way to verify that is 
by exploiting
the one-to-one correspondence between the terms thusly generated and the
Cutkosky cuts of the BFMFG two-loop self-energy;
the latter are infrared finite
since they effectively originate from a single
logarithm. 

In conclusion, we have shown that the same physical principles,
and, evidently, the same procedure used at one-loop, lead to
the generalization of the PT to two-loops. In particular , the
known correspondence between PT and BFMFG persists. It would be
interesting to explore its origin further,
and establish a formal, non-diagrammatic understanding of the PT.

\end{document}